\documentclass[epj]{svjour}
\usepackage{amsfonts}
\usepackage{amsmath}
\usepackage{amssymb,bbold}
\usepackage{kantlipsum,widetext}
\RequirePackage{graphicx}
\RequirePackage{mathptmx}
\RequirePackage{flushend}
\RequirePackage[colorlinks,citecolor=blue,urlcolor=blue,linkcolor=blue]{hyperref}
\RequirePackage[numbers,sort&compress]{natbib}
\RequirePackage{color}

\journalname{Eur. Phys. J. Plus}
\begin{document}

\title{Proper treatment of scalar and vector exponential potentials in the Klein-Gordon equation: Scattering and bound states
}

\titlerunning{Proper treatment of scalar and vector exponential potentials in the Klein-Gordon equation}        

\author{Elvis J. Aquino Curi\inst{1}\and Luis B. Castro\inst{1,2}\and Antonio S. de Castro\inst{2}
}

\institute{Departamento de F\'{\i}sica, Universidade Federal do Maranh\~{a}o (UFMA), Campus Universit\'{a}rio do Bacanga, 65080-805, S\~{a}o Lu\'{\i}s, MA, Brazil, \email{lrb.castro@ufma.br,luis.castro@pq.cnpq.br} \and
Departamento de F\'{\i}sica e Qu\'{\i}mica, Universidade Estadual Paulista (UNESP), Campus de Guaratin\-gue\-t\'{a}, 12516-410, Guaratinguet\'{a}, SP, Brazil, \email{castro@pq.cnpq.br}
}

\date{Received: date / Accepted: date}

\abstract{
We point out a misleading treatment in the literature regarding to bound-state solutions for the $s$-wave Klein-Gordon equation with exponential scalar and vector potentials. Following the appropriate procedure for an arbitrary mixing of scalar and vector couplings, we generalize earlier works and present the correct solution to bound states and additionally we address the issue of scattering states. Moreover, we present a new effect related to the polarization of the charge density in the presence of weak short-range exponential scalar and vector potentials.
\PACS{03.65.Pm \and 11.80.-m \and 03.65.Ge}
}

\maketitle

\section{Introduction}
\label{intro}

The solution of the Klein-Gordon (KG) equation with the exponential
potential may find applications in the study of pionic atoms, doped Mott
insulators, doped semiconductors, interaction between ions, quantum dots
surrounded by a dielectric or a conducting medium, protein structures, etc.
Bound-state and scattering $s$-wave solutions of the three-dimensional (KG)
equation for a minimally coupled exponential potential have already been
analyzed in the literature \cite{INC23:311:1974}. The bound states were
revisited in \cite{ZNA39:703:1984}, and the continuum in \cite{ZPD8:315:1988}%
. The authors of \cite{ZNA39:703:1984}, though, were not able to reproduce
the numerical results found in \cite{INC23:311:1974}, and the solution for
continuum states found in \cite{ZPD8:315:1988} differs from that one found
in \cite{INC23:311:1974}. Later, bound-state $s$-wave solutions received
attention for a mixing of vector and scalar couplings with different
magnitudes \cite{PLA339:300:2005,PLA349:87:2006,IJTP47:1612:2008}. The authors of \cite{PLA339:300:2005} and \cite%
{PLA349:87:2006} used a quantization condition founded on a wrong boundary
condition on the radial eigenfunction at the origin by considering the limit
to infinity of a variable necessarily finite, so turning Kummer's function into a polynomial. In \cite{IJTP47:1612:2008}, a method to solve
Kummer's equation was applied without paying attention to the proper behaviour of
the radial eigenfunction at the origin, obtaining in that way a polynomial
expression to Kummer's function. In Ref. \cite{Hassanabadi2011}, the authors addressed the
problem for arbitrary angular momentum in $D$ dimensions with arbitrary
scalar-vector mixing plus an exponential position dependent mass. A position
dependent mass can be seen as an additional scalar potential in the KG
theory. Manifestly, the eigenfunctions found in Ref. \cite{Hassanabadi2011}
satisfy a wrong boundary condition at the origin. It is worthwhile to
mention that bound-state solutions for the symmetric exponential potential
in the one-dimensional case (sometimes called cusp potential or screened
Coulomb potential) have also received attention for vector \cite%
{PLA296:192:2002,PLA362:21:2007}, scalar \cite{IJMPA17:4793:2002}
and a general mixing of vector and scalar \cite{IJMPA21:5141:2006}
couplings. Scattering in a repulsive exponential potential minimally coupled
has been studied in \cite{IJMPA21:313:2006}. For an enough deep and narrow
vector potential it might appear additional antiparticle bound states in a potential attractive
only for particles, the phenomenon called
Schiff-Snyder-Weinberg (SSW) effect \cite{PR57:315:1940}.

This work presents a detailed qualitative and quantitative analyses of
continuum $s$-wave solutions of the KG equation for attractive or repulsive
exponential potentials with arbitrary mixing of vector and scalar couplings
in a three-dimensional space. Quantization condition and constraints on the
potential parameters for bound states are identified by two different
processes: vanishing of the radial eigenfunction at infinity and poles of
the scattering amplitude. With this systematic plan of action we not only
generalize previous approaches but also elucidate some important obscure
points referred to in the previous paragraph. A particular case of our
results for the continuum gives support to that found in \cite{ZPD8:315:1988}%
. Unquestionable bound states satisfying proper boundary conditions at the
origin and at infinity are obtained from the zeros of Bessel's function of
the first kind in the case of vector and scalar couplings with equal
magnitudes, or from the zeros of Kummer's function (confluent hypergeometric
function) in the case of vector and scalar couplings with unequal
magnitudes. There is no need for breaking off the series defining Bessel's
function or Kummer's function.

\section{KG equation with vector and scalar interactions}
\label{sec:1}

The time-independent KG equation for a spinless particle with rest mass $m$
and energy $E$ under the influence of external scalar, $S$, and vector, $V$,
interactions reads ($\hbar =c=1$) 
\begin{equation}
\left[ \nabla ^{2}+\left( E-V\right) ^{2}-\left( m+S\right) ^{2}\right] \phi
=0,  \label{eqkg}
\end{equation}%
\noindent with charge density and charge current density expressed as%
\begin{equation}
\varrho =\frac{E-V}{m}|\phi |^{2},\quad \vec{J}=\frac{i}{2m}\left( \phi 
\overrightarrow{\nabla }\phi ^{\ast }-\phi ^{\ast }\overrightarrow{\nabla }%
\phi \right) .  \label{rj}
\end{equation}%
\noindent Note that if $\phi $ is a solution for a particle (antiparticle)
with energy $E$ for the potentials $V$ and $S$, then $\pm \phi ^{\ast }$ is
a solution for a antiparticle (particle) with energy $-E$ and for the
potentials $-V$ and $S$. It is also valuable to note that one finds the
nonrelativistic regime governed by Schr\"{o}dinger equation%
\begin{equation}
\left[ \nabla ^{2}+2m\left( \pm E-m-S\mp V\right) \right] \phi =0,
\end{equation}%
\noindent for weak couplings and $E\simeq \pm m$.  In the
nonrelativistic regime one finds $\varrho \simeq \pm |\phi |^{2}$ for $%
E\simeq \pm m$. However, the charge density has not a definite sign for
strong vector couplings. Of course, the resulting binding force depends on
the average charge closer to the center of force. Therefore, not only for
strong couplings, intrinsically relativistic effects can also be related to
short-range vector potentials due the polarization of the charge density.

For spherically symmetric interactions, i.e. $S(\overrightarrow{r})=S(r)$ and 
$V(\overrightarrow{r})=V(r)$, the wave function can be factorized as 
\begin{equation}
\phi _{\mu lm_{l}}(\overrightarrow{r})=\frac{u_{\mu }(r)}{r}Y_{lm_{l}}(\theta ,\varphi
)\,,  \label{eqwf}
\end{equation}%
\noindent where $Y_{lm_{l}}(\theta ,\varphi )$ is the usual spherical
harmonic, with $l=0,1,2,\ldots $, $m_{l}=-l,-l+1,\ldots ,l$ and $\mu $
denotes the principal quantum number plus other possible quantum numbers
which may be necessary to characterize $\phi $.

The radial function $u(r)$ obeys the radial equation (for $l=0$, $s$-wave) 
\begin{equation}
\frac{d^{2}u}{dr^{2}}+\left[ k^{2}+V^{2}-S^{2}-2(EV+mS)\right] u=0\,,
\label{equ}
\end{equation}%
\noindent with $k=\sqrt{E^{2}-m^{2}}$. Eq. (\ref{equ}) is effectively the
time-independent Schr\"{o}dinger equation with the effective potential $%
S^{2}-V^{2}+2(EV+mS)$. One can see that the effective potential tends to $%
S^{2}-V^{2}$ for potentials which tend to infinity at large distances so
that the KG equation furnishes a purely discrete (continuous) spectrum for $%
|S|>|V|$ ($|S|<|V|$). On the other hand, if the potentials vanish at large
distances the continuum spectrum is omnipresent but the necessary conditions
for the existence of a discrete spectrum is not an easy task for general
functional forms. Assuming that $r^{2}S(r)$ and $r^{2}V(r)$ go to zero as $r\rightarrow0$, one must
impose the homogeneous Dirichlet condition $u(0)=0$ (see, e.g. \cite%
{BAYM1990}). On the other hand, if both potentials vanish at large distances
the solution $u$ behaves like $e^{\pm ikr}$ as $r\rightarrow
\infty $. 

For scattering states in spherically symmetric scatterers, the scattering
amplitude can be written as a partial wave series (see, e.g. \cite{BAYM1990})%
\begin{equation}
f_{k}\left( \theta \right) =\sum\limits_{l=0}^{\infty }\left( 2l+1\right)
f_{l}\left( k\right) P_{l}\left( \cos \theta \right) ,  \label{f}
\end{equation}%
\noindent where $\theta $ is the angle of scattering, $P_{l}$ is the Legendre
polynomial of order $l$ and the partial scattering amplitude is%
\begin{equation}
f_{l}\left( k\right) =\left[ e^{2i\delta _{l}\left( k\right) }-1\right]
/\left( 2ik\right) .  \label{fl}
\end{equation}%
\noindent For elastic scattering the phase shift $\delta _{l}\left( k\right) $ is a
real number in such a way that at large distances%
\begin{equation}
u(r)\sim e^{-ikr}+\left( -1\right) ^{l+1}e^{2i\delta _{l}\left( k\right)
}e^{+ikr}.
\end{equation}%
\noindent Information about the energies of the bound-state solutions can be obtained
from poles of the partial scattering amplitude when one considers $k$
imaginary, but it carries an important caveat: not all the poles correspond
to bound states. For potentials with range $a$ one finds $ka\ll l$ (see,
e.g. \cite{BAYM1990}). Hence, for short-range potentials and low enough
momentum the partial wave series converges rapidly and the contribution is
predominantly $s$-wave, i.e. $f_{k}\left( \theta \right) \approx f_{0}\left(
k\right) $, which is of great importance for what follows.

\section{Exponential potentials}

Let us consider scalar and vector exponential interactions in the form 
\begin{equation}
S(r)=-S_{0}e^{-\alpha r}\,,\quad V(r)=-V_{0}e^{-\alpha r}\,,  \label{eqpot}
\end{equation}%
\noindent where $\alpha $ is a positive constant. Substituting (\ref{eqpot})
into (\ref{equ}) we get 
\begin{equation}
\frac{d^{2}u}{dr^{2}}+\left( k^{2}-V_{1}e^{-\alpha r}-V_{2}e^{-2\alpha
r}\right) u=0,  \label{eqef}
\end{equation}%
\noindent where 
\begin{equation}
V_{1}=-2(EV_{0}+mS_{0})\,,\quad V_{2}=S_{0}^{2}-V_{0}^{2}\,.  \label{eqv1}
\end{equation}%
\noindent Eq. (\ref{eqef}) is effectively the time-independent Schr\"{o}%
dinger equation for the exponential potential when $|S_{0}|=|V_{0}|$, and
for the generalized Morse potential when $|S_{0}|\neq |V_{0}|$. These
effective potentials have well structures when $V_{1}<0$ and $V_{2}>0$, or $%
V_{1}<-V_{2}$ and $V_{2}<0$. Bound states are expected for $|E|<m$. By the
way, the positive(negative)-energy solutions are not to be promptly
identified with the bound states for particles (antiparticles). Rather,
whether it is positive or negative, an eigenenergy can be unambiguously
identified with a bound-state solution for a particle (antiparticle) only by
observing if the energy level emerges from the upper (lower) continuum.

When $|S_{0}|=|V_{0}|$, no bound state is expected when $S_{0}<0$.
Nevertheless, when $S_{0}>0$ and $V_{0}=+$ $S_{0}$ ($V_{0}=-$ $S_{0}$) the
well potential is deeper (shallower) for positive-energy levels than that
one for negative-energy levels, and bound states with $E\simeq -m$ ($E\simeq
+m$) can only be found asymptotically as $S_{0}$ increases. In this
particular case, one can asseverate that the discrete spectrum consists only
of particle (antiparticle) energy levels with no chance for pair production
associate with Klein's paradox.

When $|S_{0}|\neq |V_{0}|$ the possible existence of bound-state solutions
permits us to distinguish two subclasses: a) $V_{1}<0$ and $V_{2}>0$,
corresponding to $S_{0}+V_{0}E/m>0$ with $S_{0}>|V_{0}|$, allowing the
presence of energy levels with $E\simeq \pm m$; b) $V_{1}<-V_{2}$ and $%
V_{2}<0$, with positive(negative)-energy levels occurring exclusively for $%
V_{0}>0$ ($V_{0}<0$) with $-|V_{0}|<S_{0}<m-\sqrt{m^{2}+V_{0}^{2}}$. More
than this, there may appear energy levels for $V_{0}\gtrless 0$ with $%
E\simeq \pm m$ when $|V_{0}|<m$ and $|S_{0}|<|V_{0}|$, and $E\simeq \mp m$
when $|V_{0}|>m$ and $-|V_{0}|+2m<S_{0}<|V_{0}|$. In this last case, the
spectrum including $E\simeq \mp m$ for a strong pure vector coupling with $%
V_{0}\gtrless \pm 2m$ may be related either to the SSW effect or to energy
levels of particles (antiparticles) diving into the continuum of
antiparticles (particles). Because the scalar coupling does not contribute
to the polarization of the charge density its addition contributes to lower
the threshold of this peculiar effect: $V_{0}=\pm m$ when $S_{0}=m$.

Now we move to consider a quantitative treatment of our problem by
considering the two distinct classes of effective potentials.

\section{The effective exponential potential ($|S_{0}|=|V_{0}|$)}

With the change of variable%
\begin{equation}
y=y_{0}e^{-\alpha r/2}  \label{eq1}
\end{equation}%
\noindent and the definitions%
\begin{equation}
y_{0}=\frac{2i\sqrt{V_{1}}}{\alpha },\quad \nu =\frac{2ik}{\alpha },
\label{eq2}
\end{equation}%
\noindent Eq. (\ref{eqef}) becomes Bessel's equation of order $\nu $%
\begin{equation}
y^{2}\frac{d^{2}u}{dy^{2}}+y\frac{du}{dy}+\left( y^{2}-\nu ^{2}\right) u=0.
\label{eq3}
\end{equation}

One solution of this equation is Bessel's function of the first kind of
order $\nu $ \cite{FRANK2010}%
\begin{equation}
J_{\nu }\left( y\right) =\sum\limits_{j=0}^{\infty }\left( -1\right) ^{j}%
\frac{\left( y/2\right) ^{\nu +2j}}{j!\Gamma \left( \nu +j+1\right) },
\label{bes}
\end{equation}%
\noindent where $\Gamma \left( z\right) $ denotes the meromorphic gamma function with
no zeros, and with simple poles $z=0,-1,-2,\ldots $ Bessel's function $%
J_{\nu }\left( z\right) $ is an analytic function, except for a branch point
at $z=0$. The principal branch of $J_{\nu }\left( z\right) $ is analytic in
the $z$-plane cut along the interval $(-\infty ,0]$. For $z\neq 0$ each
branch of $J_{\nu }\left( y\right) $ is entire in $\nu $. Bessel's function
of real order has an infinite number of positive zeros $j_{\nu n}$, where $n$
designates the $n$-th zero, and all of these zeros are simple. The zeros
obey the inequalities $j_{\nu ,n}<$ $j_{\nu +1,n}\,$ $j_{\nu ,n+1}$ when $%
\nu \geqslant 0$ \cite{FRANK2010}.

The general solution of Eq. (\ref{eq3}) can be expressed as%
\begin{equation}
u(y)=AJ_{+\nu }\left( y\right) +BJ_{-\nu }\left( y\right) ,\quad \nu \neq 
\mathrm{integer}.  \label{eq4}
\end{equation}%
\noindent The condition $\left. u\right\vert _{r=0}=0$ makes%
\begin{equation}
u(y)=\left\{ 
\begin{array}{cc}
AJ_{-\nu }\left( y\right) , & \mathrm{for\quad }J_{-\nu }\left( y_{0}\right)
=0, \\ 
&  \\ 
A\left[ J_{+\nu }\left( y\right) -\frac{J_{+\nu }\left( y_{0}\right) }{%
J_{-\nu }\left( y_{0}\right) }J_{-\nu }\left( y\right) \right] , & \mathrm{%
for\quad }J_{-\nu }\left( y_{0}\right) \neq 0,%
\end{array}%
\right.  \label{eq5}
\end{equation}%
\noindent and the limiting form for small argument of Bessel' function prescribes that 
$u$ behaves for large $r$ as%
\begin{widetext}
\begin{equation}
u(r)\sim \left\{ 
\begin{array}{cc}
\frac{\left( y_{0}/2\right) ^{-\nu }}{\Gamma \left( 1-\nu \right) }e^{+\nu
\alpha r/2}, & \mathrm{for\quad }J_{-\nu }\left( y_{0}\right) =0, \\ 
&  \\ 
\frac{\left( y_{0}/2\right) ^{\nu }}{\Gamma \left( 1+\nu \right) }e^{-\nu
\alpha r/2}-\frac{\left( y_{0}/2\right) ^{-\nu }}{\Gamma \left( 1-\nu
\right) }\frac{J_{+\nu }\left( y_{0}\right) }{J_{-\nu }\left( y_{0}\right) }%
e^{+\nu \alpha r/2}, & \mathrm{for\quad }J_{-\nu }\left( y_{0}\right) \neq 0.%
\end{array}%
\right.  \label{eq7}
\end{equation}%
\end{widetext}
\noindent For $J_{-\nu }\left( y_{0}\right) =0$, the asymptotic behaviour only
suggests that bound states might exist if $\nu <0$. As for $J_{-\nu }\left(
y_{0}\right) \neq 0$, the asymptotic behaviour suggests that bound states
might exist if $\nu >0$ and $J_{+\nu }\left( y_{0}\right) =0$, and
scattering states requires that $\nu $ is an imaginary.

\subsection{Bound states}

In this case $k$ is an imaginary number. This means that $|E|<m$. Regardless
the sign of $\nu $ and explicitly using the fact that $\nu $ is not an
integer, the condition determining bound-state solutions takes the concise
form%
\begin{equation}
J_{2|k|/\alpha }\left( y_{0}\right) =0,  \label{eq8}
\end{equation}%
\noindent with corresponding eigenfunction expressed as%
\begin{equation}
u(r)=AJ_{2|k|/\alpha }\left( y_{0}e^{-\alpha r/2}\right) .  \label{eq8b}
\end{equation}%
\noindent Because of the way the zeros of Bessel's function of positive order
interlace, one conclude that the $s$-wave spectrum is nondegenerate. The
order of Bessel's function in Eq. (\ref{eq8}) is a positive number so that $%
y_{0}>0$ and the effective exponential potential has a well structure when $%
S_{0}>0$. As a result from Eqs. (\ref{eqv1}) and (\ref{eq2}), $S_{0}$ must
be enough strong to make the existence of bound states possible. In fact,
Eq. (\ref{eq8}) has at least one solution when 
\begin{equation}
S_{0}>\frac{\left( \alpha j_{2|k|/\alpha ,1}\right) ^{2}}{8\left( m\pm
E\right) },\quad V_{0}=\pm S_{0}.
\end{equation}%
\noindent Therefore, one conclude that a solution with $E\simeq \pm m$ appears if $%
S_{0}\gtrsim \left( \alpha j_{0,1}\right) ^{2}/(16m)$. Consequently, bound
states in a weak potential are only allowed if the range of the potential is
enough large. On the other hand, solutions with $E\simeq \mp m$ might appear
for very large $S_{0}$. Bound-state solutions in a short-range potential
need strong couplings.

\subsection{Scattering states}

As for the continuous spectrum, $k$ is a real number so that $|E|>m$. From
the second line of Eq. (\ref{eq7}), the asymptotic form of $u$ for large $r$
clearly shows incoming and outgoing partial $s$-waves with amplitudes
differing by factors related to the phase shift: 
\begin{equation}
u(r)\sim \frac{\left( y_{0}/2\right) ^{2ik/\alpha }}{\Gamma \left(
1+2ik/\alpha \right) }e^{-ikr}-\frac{\left( y_{0}/2\right) ^{-2ik/\alpha }}{%
\Gamma \left( 1-2ik/\alpha \right) }\frac{J_{+2ik/\alpha }\left(
y_{0}\right) }{J_{-2ik/\alpha }\left( y_{0}\right) }e^{+ikr},  \label{eq9}
\end{equation}%
\noindent in such a way that%
\begin{equation}
e^{2i\delta _{0}}=\left( \frac{2}{y_{0}}\right) ^{4ik/\alpha }\frac{\Gamma
\left( 1+2ik/\alpha \right) }{\Gamma \left( 1-2ik/\alpha \right) }\frac{%
J_{+2ik/\alpha }\left( y_{0}\right) }{J_{-2ik/\alpha }\left( y_{0}\right) }.
\end{equation}%
\noindent Therefore, if one considers the analytic continuation for the entire complex 
$k$-plane $J_{+2ik/\alpha }\left( y_{0}\right) $ is an analytic function of $%
k$, and the same happens with $\Gamma \left( 1+2ik/\alpha \right) $ except
for $k=i\alpha n/2$ with $n=1,2,3\ldots $ Hence, the partial scattering
amplitude for $s$-waves is analytical in the entire complex $k$-plane,
except for isolated singularities related either to the poles of the gamma
function or to the zeros of $J_{-2ik/\alpha }\left( y_{0}\right) $. It is
true that poles of the partial scattering amplitude for $s$-waves make 
\begin{equation}
u(r)\sim e^{-|k|r},\quad kr\gg 1,
\end{equation}%
\noindent so that they could be related to bound states. However, poles of the gamma
function do not furnish licit bound states because they make $\nu
=-1,-2,-3,\ldots $, values already excluded from the general solution
expressed by Eq. (\ref{eq4}). There remains

\begin{equation}
J_{-2ik/\alpha }\left( y_{0}\right) =0.
\end{equation}%
\noindent As seen before, only the solution with $k=+i|k|$ correspond to bound states.

\section{The effective generalized Morse potential ($|S_{0}|\neq |V_{0}|$)}

With the changes%
\begin{equation}
y=y_{0}e^{-\alpha r},\quad u(y)=y^{-1/2}w(y)
\end{equation}%
\noindent and the definitions%
\begin{equation}
y_{0}=\frac{2\sqrt{V_{2}}}{\alpha },\quad \kappa =-\frac{V_{1}}{\alpha
^{2}y_{0}}\,,\quad \nu =\frac{ik}{\alpha },
\end{equation}%
\noindent Eq. (\ref{eqef}) becomes Whittaker's equation 
\begin{equation}
\frac{d^{2}w}{dy^{2}}+\left( -\frac{1}{4}+\frac{\kappa }{y}+\frac{1/4-\nu
^{2}}{y^{2}}\right) w=0.  \label{whi}
\end{equation}%
\noindent

\bigskip The general solution of (\ref{whi}) can be expressed in terms of
Kummer's function \cite{FRANK2010} 
\begin{equation}
M(a_{1},b_{1},z)=\frac{\Gamma (b_{1})}{\Gamma (a_{1})}\sum_{n=0}^{\infty }%
\frac{\Gamma (a_{1}+n)}{\Gamma (b_{1}+n)}\frac{z^{n}}{n!},\quad b_{1}\neq
0,-1,-2,\ldots  \label{conf}
\end{equation}%
\noindent Kummer's function $M(a_{1},b_{1},z)$ is entire in $z$ and $a_{1}$, and is a
meromorphic function of $b_{1}$ with simple poles at $b_{1}=0,-1,-2,\ldots $
It converges to $e^{z}z^{a_{1}-b_{1}}/\Gamma \left( a_{1}\right) $ as $%
z\rightarrow \infty $ and has an infinite set of complex zeros when $a_{1}$
and $b_{1}-a_{1}$ are different from $0,-1,-2,\ldots $ The number of real
zeros is finite when both $a_{1}$ and $b_{1}$ are real. For $b_{1}\geq 0$,
Kummer's function has no zeros when $a_{1}\geqslant 0$, and a number of
positive zeros given by the ceiling of $-a_{1}$ when $a_{1}<0$ \cite%
{FRANK2010}.

Now, $y$ lies in the interval $(0,y_{0}]$ and the general solution of Eq. (%
\ref{whi}) is expressed as%
\begin{equation}
w\left( y\right) =y^{1/2}e^{-y/2}\left[ Ay^{\nu }M^{(+)}(y)+By^{-\nu
}M^{(-)}(y)\right]
\end{equation}%
\noindent with%
\begin{equation}
M^{(\pm )}(y)=M(1/2-\kappa \pm \nu ,1\pm 2\nu ,y).
\end{equation}%
\noindent Therefore,%
\begin{equation}
u\left( y\right) =e^{-y/2}\left[ Ay^{\nu }M^{(+)}(y)+By^{-\nu }M^{(-)}(y)%
\right] .  \label{uuu}
\end{equation}%
\noindent The condition $\left. u\right\vert _{r=0}=0$ enforces%
\begin{widetext}
\begin{equation}
u(y)=\left\{ 
\begin{array}{cc}
Ae^{-y/2}y^{-\nu }M^{(-)}(y), & \mathrm{for\quad }M^{(-)}(y_{0})=0, \\ 
&  \\ 
Ae^{-y/2}\left[ \left( \frac{y}{y_{0}}\right) ^{\nu }M^{(+)}(y)-\frac{%
M^{(+)}(y_{0})}{M^{(-)}(y_{0})}\left( \frac{y}{y_{0}}\right) ^{-\nu
}M^{(-)}(y)\right] , & \mathrm{for\quad }M^{(-)}(y_{0})\neq 0.%
\end{array}%
\right.
\end{equation}%
\end{widetext}
\noindent From Eq. (\ref{conf}), $M(a_{1},b_{1},0)=1$, hence one gets the
following asymptotic expression for $u$ at large distance: 
\begin{equation}
u\left( r\right) \sim \left\{ 
\begin{array}{cc}
y_{0}^{-\nu }e^{+\nu \alpha r}, & \mathrm{for\quad }M^{(-)}(y_{0})=0, \\ 
&  \\ 
e^{-\nu \alpha r}-\frac{M^{(+)}(y_{0})}{M^{(-)}(y_{0})}e^{+\nu \alpha r}, & 
\mathrm{for\quad }M^{(-)}(y_{0})\neq 0.%
\end{array}%
\right.  \label{34}
\end{equation}%
\noindent For $M^{(-)}(y_{0})=0$, the asymptotic behaviour only suggests that bound
states might exist if $\nu <0$. As for $M^{(-)}(y_{0})\neq 0$, the
asymptotic behaviour suggests that bound states might exist if $\nu >0$ and $%
M^{(+)}(y_{0})=0$, and scattering states requires that $\nu $ is an
imaginary number.

\subsection{Bound states}

In this case $k$ is imaginary ($|E|<m$). The quantization condition takes
the concise form%
\begin{equation}
M(1/2-\kappa +|k|/\alpha ,1+2|k|/\alpha ,y_{0})=0,  \label{34a}
\end{equation}%
\noindent with corresponding eigenfunction expressed as%
\begin{equation}
u(r)=Ay_{0}^{|k|/\alpha }e^{-|k|r-y_{0}e^{-\alpha r/2}}M(1/2-\kappa
+|k|/\alpha ,1+2|k|/\alpha ,y_{0}e^{-\alpha r}).  \label{34b}
\end{equation}
\noindent $|S_{0}|>|V_{0}|$ makes $V_{2}>0$, and the relation mentioned before
involving the number of positive zeros and the parameters of Kummer's
function requires $V_{1}<0$. This is actually when the effective generalized
Morse potential has a well structure. Furthermore, $S_{0}>|V_{0}|$, and%
\begin{equation}
|V_{1}|/\sqrt{V_{2}}\gtrsim \alpha .  \label{cond}
\end{equation}%
\noindent This last condition let us to get some conclusions regarding the range of
the potential. One finds $E\simeq \pm m$ for%
\begin{equation}
S_{0}\gtrsim \mp V_{0}\frac{1+\beta }{1-\beta },\quad \beta =\left( \frac{%
\alpha }{2m}\right) ^{2}.
\end{equation}%
\noindent Hence, solutions with $E\simeq \pm m$ ($E\simeq \mp m$) and $V_{0}\gtrless 0$
for large(short)-range potentials just demand $S_{0}>|V_{0}|$, whereas
solutions with $E\simeq \mp m$ ($E\simeq \pm m$) demand a stronger scalar
coupling $S_{0}\gtrsim |V_{0}|\left( 1+2\beta \right) $ ($S_{0}\gtrsim
|V_{0}|\left( 1+2/\beta \right) $). Even for weak potentials, the absence of
solutions with $E\simeq \pm m$ for $|V_{0}|<S_{0}\lesssim |V_{0}|\left(
1+2/\beta \right) $ for short-range potentials is a genuine relativistic
quantum effect. Due to the polarization of the charge density, an attractive
(repulsive) vector coupling for particles (antiparticles) for a large-range
potential undergoes reversion of its effects as the range of the potential
decreases. Because the stronger scalar coupling is always attractive, the
final outcome for this sort of mixing of couplings is that the
large(short)-range potential is more attractive for particles
(antiparticles).

As for $|S_{0}|<|V_{0}|$, one obtains the set of imaginary zeros.
Unfortunately, due to the lack of necessary information about the set of
imaginary zeros of Kummer's function, we can not say more than we have
already said before.

It does not take long to convince oneself that one can find bound-state
solutions for $|S_{0}|<|V_{0}|$ as well as for $|S_{0}|>|V_{0}|$, at least
for short-range strong potentials. In this case, with $S_{0}$ and $V_{0}$
proportional to $\alpha $, the quantization condition (\ref{34a}) can be
approximate by $M(1/2,1,y_{0})=0$. Relations between Kummer's function with $%
a_{1}=1/2$ and $b_{1}=1$ and Bessel's function, viz. $M(1/2,1,2z)=e^{z}J_{0}%
\left( z\right) $ and $M(1/2,1,2iz)=e^{iz}J_{0}\left( z\right) $ (see, e.g. 
\cite{FRANK2010}), make Eq. (\ref{34a}) equivalent to $J_{0}\left(
|y_{0}|/2\right) =0$ in such way that one finds at least one solution when $%
|S_{0}^{2}-V_{0}^{2}|\gtrsim \left( \alpha j_{0,1}\right) ^{2}$.

\subsection{Scattering states}

As for the continuous spectrum, $k$ is a real number so that $|E|>m$. From
the second line of Eq. (\ref{34}), the asymptotic form of $u$ for large $r$
clearly shows incoming and outgoing partial $s$-waves with amplitudes
differing by factors related to the phase shift: 
\begin{equation}
u\left( r\right) \sim e^{-ikr}-\frac{M^{(+)}(y_{0})}{M^{(-)}(y_{0})}e^{+ikr},
\end{equation}%
\noindent in such a way that%
\begin{equation}
e^{2i\delta _{0}}=\frac{M(1/2-\kappa +ik/\alpha ,1+2ik/\alpha ,y_{0})}{%
M(1/2-\kappa -ik/\alpha ,1-2ik/\alpha ,y_{0})}.
\end{equation}%
\noindent Therefore, if one considers the analytic continuation for the entire complex 
$k$-plane $M(1/2-\kappa +ik/\alpha ,1+2ik/\alpha ,y_{0})$ is an analytic
function of $k$, except for $k=i\alpha n/2$ where $n$ is a nonnegative
integer. Hence, the partial scattering amplitude for $s$-waves is analytical
in the entire complex $k$-plane, except for isolated singularities related
either to the poles of $M(1/2-\kappa +ik/\alpha ,1+2ik/\alpha ,y_{0})$ or to
the zeros of $M(1/2-\kappa -ik/\alpha ,1-2ik/\alpha ,y_{0})$. It is true
that poles of the partial scattering amplitude for $s$-waves make 
\begin{equation}
u(r)\sim e^{-|k|r},\quad kr\gg 1,
\end{equation}%
\noindent so that they could be related to bound states. However, poles of $%
M(1/2-\kappa +ik/\alpha ,1+2ik/\alpha ,y_{0})$ do not furnish licit bound
states because they make $\nu =-1,-2,-3,\ldots $, values already excluded
from the general solution expressed by Eq. (\ref{eq4}). There remains

\begin{equation}
M(1/2-\kappa -ik/\alpha ,1-2ik/\alpha ,y_{0})=0.  \label{p1}
\end{equation}

\section{Final remarks}

In this work, we pointed out a misleading treatment in the literature
regarding to bound-state solutions for the $s$-wave KG equation with
exponential scalar and vector potentials in a three-dimensional space. We
showed a detailed qualitative and quantitative analyses of continuum $s$%
-wave solutions of the KG equation for attractive or repulsive exponential
potentials with arbitrary mixing of vector and scalar couplings. The care
needed in applying the proper boundary conditions was emphasized. Using the
proper boundary conditions at the origin and at infinity, we found the
quantization condition and constraints on the potential parameters for bound
states. We obtained the possible energy levels from the zeros of Bessel's
function of the first kind in the case of vector and scalar couplings with
equal magnitudes, or from the zeros of Kummer's function in the case of
vector and scalar couplings with unequal magnitudes. Although the solutions
in Refs. \cite%
{PLA339:300:2005,PLA349:87:2006,IJTP47:1612:2008,Hassanabadi2011} are licit
when one considers the limiting form of Kummer's function as $%
y_{0}\rightarrow \infty $ (either for a scalar coupling much stronger
than a vector coupling or for large-range potentials), we showed that there
is no need for breaking off the series defining Kummer's function in a more
general circumstance. Never mentioned in the literature, we showed that an
effect related to the polarization of the charge density in the presence of
short-range exponential scalar and vector potentials manifests when $%
S_{0}>|V_{0}|$ even in the case of weak couplings.

We would like to point out that our results for $s$-wave bound-state solutions (eigenvalues and also eigenfunctions) are exact, whereas the scattering amplitude is approximate. The poles of the scattering amplitude furnish results coincident with the exact bound-state solutions, as it should be in a proper treatment. If one wants to treat the case of arbitrary $l$ one should appeal to approximation methods that is out of the scope of this article. It is important to notice that our results have nothing to do with the class of multiparameter exponential-type potential studied in the Schr\"{o}dinger equation in \cite{IJQCh112:2012:195} and in the KG equation with equal vector and scalar couplings in \cite{IJP89:2015:649,FBS59:2018:52}.This is so because the class of multiparameter exponential-type potential studied in \cite{IJQCh112:2012:195} and \cite{IJP89:2015:649} does not reduce to a pure exponential potential as that one studied in this paper, and that one studied in \cite{FBS59:2018:52} does so only in an approximation scheme.

\begin{acknowledgement}
This work was supported in part by means of funds provided by CNPq, Brazil, Grants No. 304743/2015-1 (PQ), No. 307932/2017-6 (PQ) and No. 422755/2018-4 (UNIVERSAL), FAPEMA, Brazil, Grant No UNIVERSAL-01220/18, S\={a}o Paulo Research Foundation (FAPESP), Grant No. 2018/20577-4 and CAPES, Brazil. 
\end{acknowledgement}

\bibliographystyle{spphys}       

\end{document}